\documentclass[twocolumn,pra,amsmath,amssymb]{revtex4-1}
\usepackage{graphicx,bm,epsfig,color}

\begin{document}

\title{Semi-analytical RWA formalism to solve Schr\"odinger equations\\ for multi-qudit systems with resonator couplings}
\author{H.W.L. Naus}
\email{rik.naus@tno.nl}
\affiliation{TNO, P.O. Box 96864, 2509 JG The Hague, The Netherlands and
QuTech, Delft University of Technology, P.O. Box 5046, 2600 GA Delft, The Netherlands}
\author{R. Versluis}
\email{richard.versluis@tno.nl}
\affiliation{TNO, P.O. Box 155, 2600 AD Delft, The Netherlands and
QuTech, Delft University of Technology, P.O. Box 5046, 2600 GA Delft, The Netherlands}

\begin{abstract}
In this study, we develop a semi-analytical framework to solve generalized Jaynes-Tavis-Cummings
Hamiltonians describing multi-qudit systems coupled via EM resonators. Besides the 
multi-level generalization we allow for an arbitrary number of resonators and/or modes, with nonidentical
couplings to the qudits. Our method is based on generic excitation-number operators which commute with the
respective Hamiltonians in the rotating wave approximation (RWA). The validity of the RWA is assessed explicitly. 
The formalism enables the study of eigenstates, eigenenergies and corresponding time evolutions of such coupled multi-qudit systems. 
The technique can be applied in cavity quantum electrodynamics and circuit quantum electrodynamics. 
It is also applicable to atomic physics, describing the coupling of a single-mode photon to an atom. 
As an example, we solve the Schr\"odinger equation for a two-qubit-one-resonator system, in principle to arbitrary high excitations. 
We also solve the Tavis-Cummings Hamiltonian in the one-excitation subspace for an arbitrary number of identical
qubits resonantly coupled to one resonator. As a final example, we calculate of the low-excitation spectrum of a coupled two-transmon system.
\end{abstract}

\date{\today}
\maketitle

\newcommand{\ketz}{|\,\mathbf{0}\rangle}
\newcommand{\keto}{|\,\mathbf{1}\rangle}
\newcommand{\braz}{\langle\mathbf{0}\,|}
\newcommand{\brao}{\langle\mathbf{1}\,|}

\newcommand{\ketnz}{|\,n, \mathbf{0}\rangle}
\newcommand{\ketno}{|\,n, \mathbf{1}\rangle}
\newcommand{\branz}{\langle n, \mathbf{0}\,|}
\newcommand{\brano}{\langle n, \mathbf{1}\,|}

\section{Introduction}
Scalable quantum processing eventually using a large number
of qubits \cite{NC} and enabling fault-tolerant quantum computing is a topical subject; see
\cite{Mavro,Chen,Brien,Versluis} and references therein. 
In order to perform the required two-qubit gates or multi-qubit gates to reach computational
universality \cite{Lloyd,Deutsch} the qubits need to interact, e.g. using electromagnetic (EM) resonators.
Examples include transmon, Xmon and fluxmon qubits with fixed \cite{Carlo,Cor,Bar,Quin} or tuneable \cite{Chen} 
couplings as well as spin qubits connected to resonators \cite{Peters,Burk,Kubo}. 
To determine the dynamic behavior of multi-qubit systems, the Schr\"odinger equation governed
by the corresponding Hamiltonian needs to be solved. In the simple case of a qubit interacting with
a single resonator this is the well-known Jaynes-Cummings (JC) Hamiltonian \cite{JC} with known exact solutions. 
A more complex example is the collective interaction of multiple identical qubits equally
coupled to one single-mode resonator as described by the Tavis-Cummings (TC) Hamiltonian.
Exact and approximate solutions have been derived for the TC model \cite{Tavis,TavisI,Bogo,Vade}.

In this study, we develop a calculational framework 
solving Schr\"odinger equations for
multi-qudit systems which are coupled via EM resonators.
Besides the multi-level generalization
we allow for an arbitrary number of resonators and/or modes, with non-identical couplings to the qudits.
The semi-analytical formalism also incorporates qudits with different energy levels. 
The commonly used rotating wave approximation (RWA) is also adopted here \cite{vanK}.
We assess its applicability in perturbation theory \cite{Cohen2}. Given the RWA, the
methods are exact and semi-analytical in the sense that only finite-dimensional
matrices are eventually numerically diagonalized,
in principle limited only by the available computing and memory resources.
The technique can be applied in cavity quantum electrodynamics
and circuit quantum electrodynamics. It is also applicable to
atomic physics, describing the coupling of a single-mode photon to an atom.

Some caveats are appropriate to include at the onset.
First, it is tacitly assumed that the dimensions of the described physical
systems are small compared to the wavelengths of the considered radiation modes.
Secondly, we neither address finite temperature effects nor imperfect
cavities. Only ideal, closed systems are considered.
Thirdly, the qubits/qudits and the EM modes are not too far from
resonance and their coupling should not be ultrastrong - otherwise the RWA would be {\em a priori}
invalid. Finally, throughout this paper we use the concept of multiple single-mode resonators.
Our framework covers multiple EM modes in one cavity equally well; care should
be taken in the limit of an infinite number of modes \cite{Gely}.

The outline of this paper is as follows. In order to fix our notation and to introduce
the method we start by re-analysing the generic JC Hamiltonian.
Next we explicitly assess the validity of the RWA in the JC model.
In section \ref{sec:MquR}  we extend the formalism based on the excitation-number to
Hamiltonians describing multiple qubits and resonators with arbitrary couplings.
The technique is applied to a system of two qubits and one resonator in section \ref{sec:TwoQ-R}.
For this specific problem we show that the applicability extends to arbitrary high excitations.
Section \ref{sec:Reso} addresses an arbitrary number of identical qubits resonantly coupled to one resonator.
We explicity solve the Schr\" odinger equation for the one-excitation subspace.
The framework is generalized further to multi-level systems called qudits in section \ref{sec:Qudit}.
The concrete example in section \ref{sec:Trans} presents the calculation of the low-excitation
spectrum of the Hamiltonian  modeling a system consisting of two qutrits coupled to a single resonator.
Finally, in section \ref{sec:Time}, we give the formal expression for the time evolution operator
of coupled multi-qudit systems. The paper is concluded with a summary.

\section{Prelude: Jaynes-Cummings physics}
We first present the generic Jaynes-Cummings (JC) Hamiltonian \cite{JC}
describing the coupling of a single-mode photon to an atom in the
two-level approximation
\begin{equation}
   H =  - \tfrac{1}{2}\hbar\omega' \sigma_z +
   \hbar\omega(a^\dagger a+\tfrac{1}{2})
  + \hbar g (a^\dagger\sigma_++a\sigma_-),
  \label{eq:JC}
\end{equation}
where $g$ is the coupling strength. 
The photon mode, frequency $\omega$, is described by creation and annihilation operators
$a^\dagger, a$ which satisfy the canonical commutation relation $[a,a^\dagger]=1$ of a harmonic
oscillator.
We use standard Pauli matrices with spin raising and lowering operators
  $\sigma_{\pm} = \tfrac{1}{2}(\sigma_x \pm i \sigma_y)$. Note that the
  RWA has already been included. This  problem can be solved
  exactly \cite{JC,Blais,Schus,Chow}. In order to further fix our notation and
  to introduce the method which is generalized below, we present the derivation
of the solution.  
 The standard qubit states satisfy
\begin{eqnarray}
  \sigma_z \ketz = \ketz, \quad 
  \sigma_+ \ketz &=&  0,  \quad   \sigma_+ \keto =  \ketz, \nonumber\\
  \sigma_z \keto = - \keto, \quad
  \sigma_- \ketz &=&  \keto,  \quad   \sigma_- \keto =  0.
\end{eqnarray}
Unperturbated oscillator states are denoted by $|n\rangle$. The excitation-number operator
is defined as 
\begin{equation}
  \mathcal N = a^\dagger a - \tfrac{1}{2} \sigma_z,
\end{equation}
which, as can be verified by explicit calculation, commutes with the JC Hamiltonian (\ref{eq:JC}) 
$  [\mathcal N, H] = 0.$
It implies that a common set of eigenstates of $\mathcal N$ and $H$ exists.
The eigenstates of $\mathcal N$ are the product states
\begin{eqnarray}
  \mathcal N \ketnz &=&  (n-\tfrac{1}{2})  \ketnz, \nonumber\\
  \mathcal N \ketno &=&  (n+\tfrac{1}{2}) \ketno.
\end{eqnarray}
For $n=0$ we get the lowest eigenvalue of $\mathcal N$, which equals $-\tfrac{1}{2}$.
The concomitant eigenstate also satisfies
\begin{equation}
H   |\,0, \mathbf{0}\rangle = 
-\tfrac{1}{2} \hbar \Delta |\,0, \mathbf{0}\rangle,
\label{eq:GS}
\end{equation}
where $\Delta = \omega'-\omega$, cf. \cite{Blais},
and equals the ground state of the Hamiltonian.
Other eigenvalues of $\mathcal N$ are $n-\tfrac{1}{2}, n \ge 1$ with degenerate eigenstates.
In order to find the energy eigenvalues and the
excited states of the JC Hamiltonian we therefore make the {\em Ansatz}
\begin{equation}
  | E \rangle = \alpha \ketnz + \beta |\,n-1, \mathbf{1}\rangle.
  \label{eq:ansatz}
\end{equation}
For each $n \ge 1$, we obtain a two-dimensional subspace, 
  called {\it RWA strip} in \cite{Sank},
in which we have to diagonalize
the corresponding $2 \times 2$ matrix, yielding 
the energy eigenvalues
\begin{equation}
    E_n^\pm = n \hbar\omega \pm \tfrac{1}{2}\hbar
    \sqrt{4g^2n+\Delta^2}
  \label{eq:energ}
\end{equation}
and the exact orthonormal solutions
\begin{eqnarray}
  | E_n^+ \rangle &=& \alpha_n^+ \ketnz + \beta_n^+ |\,n-1, \mathbf{1}\rangle, \nonumber \\
  | E_n^- \rangle &=& \alpha_n^- \ketnz + \beta_n^- |\,n-1, \mathbf{1}\rangle.
\label{eq:ESJC}
\end{eqnarray}
The coefficients are given by
\begin{eqnarray}
  \alpha^+_n &=& \rho_n^{-1}\left(\sqrt{g^2n+\tfrac{1}{4}\Delta^2}-\tfrac{1}{2}\Delta\right) = \beta_n^-,
  \nonumber \\ \beta^+_n &=& \rho_n^{-1} g \sqrt{n} = -\alpha_n^-,
\end{eqnarray}
where  $\rho_n^2=2g^2n+\tfrac{1}{2}\Delta^2
-\Delta\sqrt{g^2n+\tfrac{1}{4}\Delta^2}$ are normalization factors.
Equivalently, the coefficients are expressed in terms of angles \cite{JC,Blais,Schus,Chow}:
\begin{eqnarray}
  \alpha^+_n = \cos \theta_n&,& \beta^+_n = \sin \theta_n \quad \text{with} \nonumber \\
  \tan{(2\theta_n)} &=& -\frac{2g\sqrt{n}}{\Delta}. 
\end{eqnarray}
The splitting in the $\pm$ energy levels (\ref{eq:energ})
corresponds to the Rabi frequencies of the two-level subsystems \cite{NC,Cohen1,Cohen1}.
The eigenstates $|E_n^\pm \rangle$ are known as
{\em dressed} states in the literature \cite{CohenT,Scully}.

\section{Validity of the RWA}
The validity of the RWA is assessed by
treating the omitted interaction terms in perturbation theory. To this end, we straightforwardly
calculate the resulting shifts in eigenenergies. 
The in the RWA omitted terms correspond to the interaction Hamiltonian 
\begin{equation}
  H_p = -\hbar g (\sigma_-a^\dagger+\sigma_+a).
  \label{eq:rwa-per}
\end{equation}
They are associated with high frequencies and are therefore usually neglected \cite{vanK}.
Alternatively, it is proposed to control these interactions \cite{Huang} or, for high
photon numbers, go beyond the RWA \cite{Sank}.
Here $H_p$ will be treated as a perturbation.  If we denote the exact energies by $\tilde E_0$ and
$\tilde E_n^\pm$, we do not obtain any modification up to first order
 $ \tilde E_0 \backsimeq  E_0, \quad
  \tilde E_n^\pm \backsimeq E_n^\pm. $
Note, however, that the eigenstates are modified in first order.
Though easily calculable, we omit the explicit expressions.
\begin{widetext}
The energy eigenvalues including second order corrections are given by
\begin{eqnarray}
  \tilde E_0 &\backsimeq&  E_0 +\hbar^2 g^2
  \left(\frac{(\beta_2^+)^2}{E_0-E_2^+}+
    \frac{(\beta_2^-)^2}{E_0-E_2^-}\right),\quad
    \tilde E_1^\pm \backsimeq  E_1^\pm + 2 \hbar^2 g^2
  \left(\frac{(\alpha_1^\pm \beta_3^+)^2}{E_1^\pm-E_3^+}+
    \frac{(\alpha_1^\pm \beta_3^-)^2}{E_1^\pm-E_3^-}\right), \nonumber\\
    \tilde E_2^\pm &\backsimeq&  E_2^\pm +\hbar^2g^2\frac{(\beta_2^\pm)^2}{E_2^\pm-E_0} 
                   + 3 \hbar^2 g^2
  \left(\frac{(\alpha_2^\pm \beta_4^+)^2}{E_2^\pm-E_4^+}+
  \frac{(\alpha_2^\pm \beta_4^-)^2}{E_2^\pm-E_4^-}\right) \\
  \text{and for $n\ge 3$} \qquad && \nonumber \\
    \tilde E_n^\pm &\backsimeq&  E_n^\pm 
    + (n+1) \hbar^2 g^2 \left(\frac{(\alpha_n^\pm \beta_{n+2}^+)^2}{E_n^\pm-E_{n+2}^+}+
  \frac{(\alpha_n^\pm \beta_{n+2}^-)^2}{E_n^\pm-E_{n+2}^-}\right) 
  + (n-1) \hbar^2 g^2 \left(\frac{(\alpha_n^\pm \beta_{n-2}^+)^2}{E_n^\pm-E_{n-2}^+}+
  \frac{(\alpha_n^\pm \beta_{n-2}^-)^2}{E_n^\pm-E_{n-2}^-}\right). \nonumber
\end{eqnarray}
Figures \ref{fi:rwa} and \ref{fi:rwa2} depict some typical results of the perturbative approach
for the four lowest excitation levels.
The qubit frequency is taken as $f'= 6$ GHz and the resonator is tuned at $f=7$ GHz.
In order to visualize the effects, the range of the coupling is extended to 10 GHz.
For couplings well below the qubit and resonator frequencies, in practice below 1 GHz, the corrections
due to omitted terms are small: less than $< 0.4\%$ for the ground state and less than $0.04\%$ for $n=1,2,3$.
These results {\em a forteriori} justify the RWA in such ranges.

\begin{figure}[htb]
\includegraphics[width=210pt]{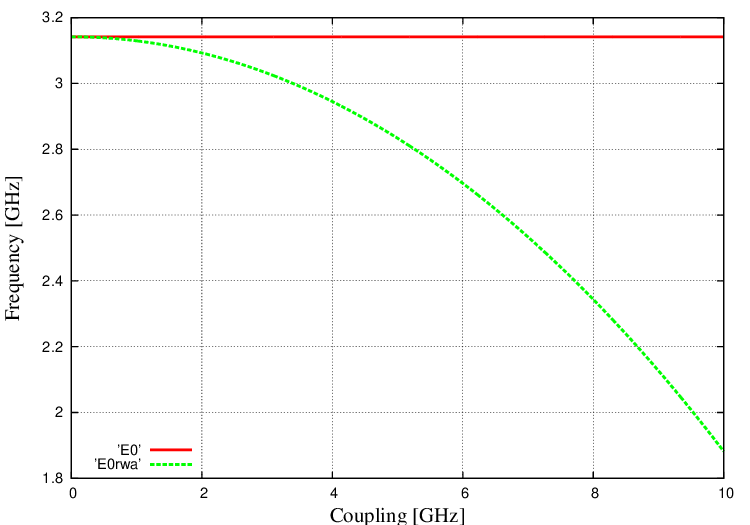}
\includegraphics[width=210pt]{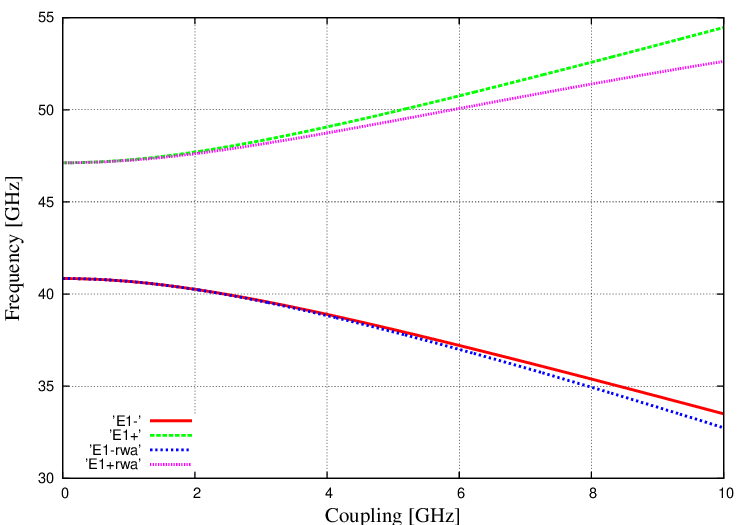}
\caption{\label{fi:rwa} Energy levels $E/\hbar$ of the ground state (l.h.s.) and first excitation (r.h.s.)
of a single qubit coupled to a single resonator as function of the coupling $g$ without and
with second order RWA correction.}
\end{figure}

\begin{figure}[htb]
\includegraphics[width=205pt]{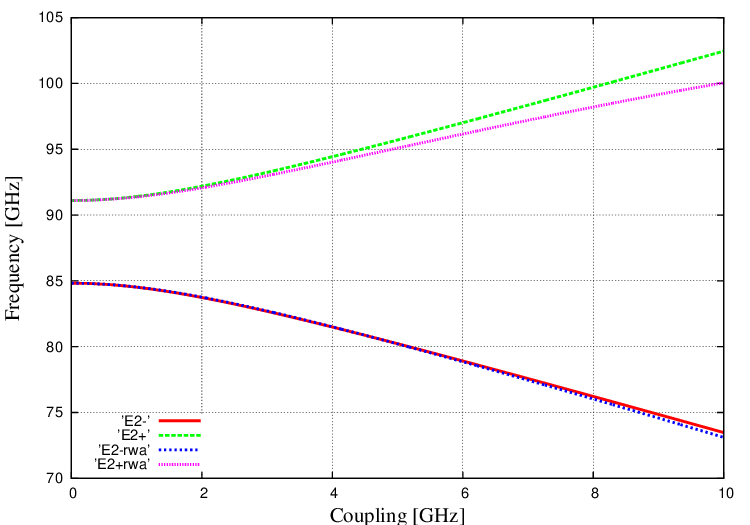}
\includegraphics[width=205pt]{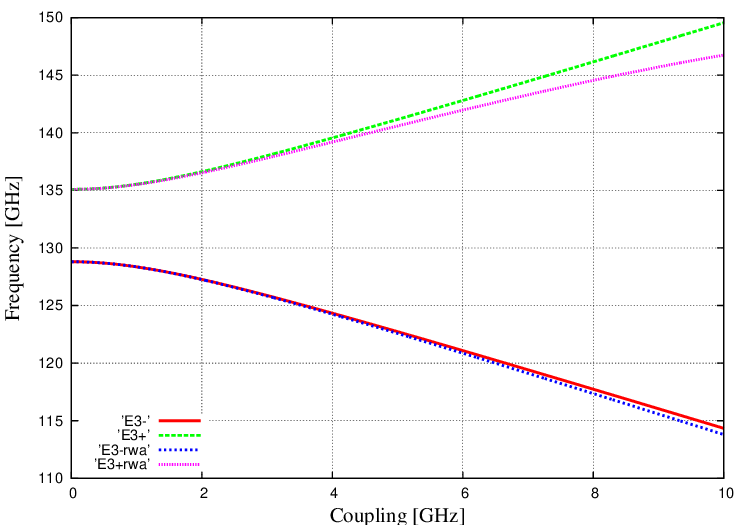}
\caption{\label{fi:rwa2}
Energy levels $E/\hbar$ of the second excitation (l.h.s.) and third excitation (r.h.s.)
of a single qubit coupled to a single resonator as function of the coupling $g$ without and
with second order RWA correction.}
\end{figure}

\end{widetext}

\section{Multiple qubits and resonators}\label{sec:MquR}
Next we consider a system of $K$ qubits and $P$ resonators. We allow the coupling of each
resonator to each qubit. 
The generalized JC Hamiltonian 
of such a system is in RWA given by
\begin{equation}
  H= \sum_{k=1}^K H_0^{[k]} + \sum_{p=1}^P H_{\text{res}}^{[p]} +\sum_{i=1}^P \sum_{j=1}^K H_{\text{int}}^{ij}.
\end{equation}
We have introduced qubit, resonator and interaction Hamiltonians respectively as
\begin{eqnarray}
  H_0^{[k]} &=&  - \tfrac{1}{2} \hbar  \omega'_k \sigma_z^{[k]}, \nonumber \\
  H_{\text{res}}^{[p]} &=&  \hbar  \omega_p(a_p^\dagger a_p+\tfrac{1}{2}), \nonumber \\
  H_{\text{int}}^{ij} &=&
  \hbar  g_{ij} \left(a_i^\dagger\sigma_+^{[j]}+a_i\sigma_-^{[j]}\right).
\end{eqnarray}
In realistic systems, many couplings out of the set $g_{ij}$ may vanish. 
The corresponding excitation-number operator $\mathcal{N}$ reads in this case 
\begin{equation}
  \mathcal{N} = \sum_{q=1}^P a_q^\dagger a_q
  - \tfrac{1}{2} \sum_{s=1}^K \sigma_z^{[s]}.
\end{equation}
In order not to overload the notation we use the {\it same} symbol
for the various excitation-number operators.
Obviously $\mathcal{N}$ commutes with the `free'  Hamiltonians
  $[ \mathcal{N}, H_0^{[k]}] = 0, \quad
  [ \mathcal{N}, H_{\text{res}}^{[p]}] = 0$.
Next we calculate the commutator of the excitation-number operator and the interaction Hamiltonians
\begin{eqnarray}
  [ \mathcal{N},  H_{\text{int}}^{ij}]
   &=& \hbar g_{ij} \left(
  \sum_{q=1}^P \delta_{qi}(a_q^\dagger \sigma_+^{[j]}- a_q \sigma_-^{[j]}) \right. \nonumber \\
 &-& \left.\tfrac{1}{2}
  \sum_{s=1}^K \delta_{sj}(2 a_i^\dagger \sigma_+^{[j]}-2a_i\sigma_-^{[j]}))\right)  \\
   &=& \hbar g_{ij} \left(
  a_i^\dagger \sigma_+^{[j]}- a_i \sigma_-^{[j]}
  - a_i^\dagger \sigma_+^{[j]}+a_i\sigma_-^{[j]}\right) = 0. \nonumber
\end{eqnarray}
As a consequence we obtain $[ \mathcal{N},  H] = 0$.
The excitation-number operator and the Hamiltonian are
therefore simultaneously diagonalizable.
To that end, we note that the product states $|n_1,\dots,n_P, \bm{b_1},\dots,\bm{b_K} \rangle$
are eigenstates of  $\mathcal{N}$ with eigenvalues $N$.
We  indicate the photon state by means of $n_p=0,1,2\cdots$ and the qubits by
$\bm{b_k=0,1}$.
The eigenvalues $N$ follow as
\begin{equation}
  N = \sum_{p=1}^P n_p + \sum_{k=1}^K (\bm{b_k}-\tfrac{1}{2})
  = \sum_{p=1}^P n_p + \sum_{k=1}^K \bm{b_k}-\tfrac{1}{2}K.
  \label{eq:eig}
\end{equation}
With the exception of the state corresponding to $N=-\tfrac{1}{2}K$, expectedly the ground state
of the Hamiltonian, the eigenstates are (highly) degenerate. The degree of degeneracy $L$
depends on $N$, {\em i.e.}, $L=L(N)$; where irrelevant we do not explicitly indicate this.
Labelling the degenerate states with the index $l$ leads to the alternative notation
$ |n_1,\dots,n_P, \bm{b_1},\dots,\bm{b_K} \rangle = |N; l \rangle.$
Of course, we expect that
the Hamiltonian does not couple the various subspaces corresponding to different values
of $N$. 
We check this by explicitly calculating its matrix elements.
The matrix elements of the `free' Hamiltonians are
\begin{equation}
  \langle N'; l'| \sum_{k=1}^K H_0^{[k]} | N; l\rangle
  = -\tfrac{1}{2} \hbar \delta_{N'N}\delta_{l'l} \sum_{k=1}^{K} (-1)^{b_k} \omega_k'
\end{equation}
and
\begin{equation}
  \langle N'; l'| \sum_{p=1}^P H_{\text{res}}^{[p]} | N; l\rangle
  =  \hbar \delta_{N'N}\delta_{l'l} \sum_{p=1}^{P} \omega_p(n_p+\tfrac{1}{2}).
\end{equation}
For the interaction terms of the Hamiltonian we get
\begin{eqnarray}
  && \langle N'; l'| \sum_{i=1}^P \sum_{j=1}^K H_{\text{int}}^{ij} | N; l\rangle \nonumber \\
  &=&  
    \hbar\sum_{i=1}^P \sum_{j=1}^K g_{ij} \delta_{n'n}^{-i}\delta_{b'b}^{-j}
  \left(\delta_{n_i'n_{i+1}}\delta_{b_j'0}\delta_{b_j1}\sqrt{n_i+1} \right. \nonumber \\
  &+& \left.\delta_{n_i'n_{i-1}}\delta_{b_j'1}\delta_{b_j0}\sqrt{n_i}\right),
  \label{eq:resi} 
\end{eqnarray}
where $\delta_{n'n}^{-i}=1$ if $n_k'=n_k$ for all $k \ne i$;
$\delta_{b'b}^{-j}$ is defined analogously. Using the Kronecker deltas it follows
that $ N = N' + n_i - n_i'+ b_j-b_j' = N'$, {\it i.e.}, 
nonzero matrix elements have equal excitation number. It is also clear that the 
Hamiltonians in the various subspaces are symmetric. 
Consequently, these matrices have real eigenvalues and
- within their particular subspace- a complete set of eigenvectors. In other words,
they are diagonalizable by the standard Jacobi procedure.
In this sense, the general multi qubit-resonator problem
is solved.

\section{Two qubits coupled by a cavity}\label{sec:TwoQ-R}
As an explicit example we first address the simple multi-qubit problem
of two qubits coupled by one electromagnetic resonator.
Thus we derive the solution of the time-independent Schr\"odinger equation
for the Hamiltonian
\begin{eqnarray}
  H &=&  - \tfrac{1}{2} \hbar \omega'_1 \sigma_z^{[1]} -
   \tfrac{1}{2} \hbar \omega'_2 \sigma_z^{[2]} +
  \hbar  \omega(a^\dagger a+\tfrac{1}{2}) \\
  &+& \hbar g_{1} \left(a^\dagger\sigma_+^{[1]}+a\sigma_-^{[1]}\right)
  + \hbar  g_{2} \left(a^\dagger\sigma_+^{[2]}+a\sigma_-^{[2]}\right). \nonumber
  \label{eq:JCq2}
\end{eqnarray}
Recall that the RWA has been made.

\subsection{Derivation of subspace Hamiltonians}
Analogously to the one-qubit problem, we define the operator
\begin{equation}
  \mathcal N = a^\dagger a - \tfrac{1}{2} \sigma_z^{[1]} - \tfrac{1}{2} \sigma_z^{[2]}.
\end{equation}
Once more, a brief calculation shows that 
\begin{equation}
  [\mathcal N, H] = 0.
\end{equation}
The vanishing commutator guarantees a common set of eigenstates of $\mathcal N$ and $H$.
The eigenstates of $\mathcal N$ are given by
\begin{eqnarray}
  \mathcal N |j, \mathbf{0, 0} \rangle  &=& (j-1) |j, \mathbf{0, 0} \rangle   \nonumber \\
  \mathcal N |k, \mathbf{1, 0} \rangle  &=& k |k, \mathbf{1, 0} \rangle   \nonumber \\
  \mathcal N |l, \mathbf{0, 1} \rangle  &=& l |l, \mathbf{0, 1} \rangle   \nonumber \\
  \mathcal N |m, \mathbf{1, 1} \rangle  &=& (m+1) |m, \mathbf{1, 1} \rangle   .
\end{eqnarray}
The lowest eigenvalue of $\mathcal N$ is -1 and the unique eigenstate is
$  |0, \mathbf{0, 0} \rangle$.
It is also an eigenstate of the Hamiltonian
(\ref{eq:JCq2}):
\begin{equation}
   H |0, \mathbf{0, 0} \rangle = 
   E_{-1} |0, \mathbf{0, 0} \rangle = 
   \tfrac{1}{2}\hbar(\omega - \omega_1'-\omega_2')
  |0, \mathbf{0, 0} \rangle,
\end{equation}
expectedly the ground state.

The next eigenvalue of $\mathcal{N}$ is equal to 0 and we see that there is a three-fold
degeneracy since the three states
  $|1, \mathbf{0, 0} \rangle,
  |0, \mathbf{1, 0} \rangle,
  |0, \mathbf{0, 1} \rangle$ correspond to this eigenvalue.
  In this subspace we therefore make the {\em Ansatz}
\begin{equation}
   | E_0 \rangle  = \alpha_0 |1, \mathbf{0, 0} \rangle  + \beta_0 |0, \mathbf{1, 0} \rangle  
   + \gamma_0 |0, \mathbf{0, 1} \rangle.
\end{equation}
The time-independent Schr\"odinger equation in this subspace gives
\begin{eqnarray}
  &\hbar& \!\!\alpha_0\left\{\left(-\tfrac{1}{2}\omega_+'+\tfrac{3}{2}\omega\right)
  |1, \mathbf{0, 0} \rangle  + g_1 |0, \mathbf{1, 0} \rangle  
+  g_2|0, \mathbf{0, 1} \rangle  \right\} \nonumber \\
&+&\hbar\beta_0\left\{\left(\tfrac{1}{2}\omega_-'+\tfrac{1}{2}\omega\right)
  |0, \mathbf{1, 0} \rangle  +  g_1 |1, \mathbf{0, 0} \rangle  \right\} \nonumber \\
  &+& \hbar\gamma_0\left\{\left(-\tfrac{1}{2}\omega_-'+\tfrac{1}{2}\omega\right)
  |0, \mathbf{0, 1} \rangle 
+  g_2 |1, \mathbf{0, 0} \rangle  \right\} \nonumber \\
  &=& E_0\left\{\alpha_0 |1, \mathbf{0, 0} \rangle  + \beta_0 |0, \mathbf{1, 0} \rangle  
   + \gamma_0 |0, \mathbf{0, 1} \rangle \right\},
\end{eqnarray}
with $\omega_\pm'=\omega_1'\pm\omega_2'$. We take the inner product with the basis states of
of this subspace and obtain  the matrix eigenvalue equation
\begin{equation}
   \mathcal{H}_0 \vec \eta_0 = E_0 \vec \eta_0, \quad \text{where} \quad  \vec\eta_0 = 
  \begin{pmatrix}
    \alpha_0 \\
    \beta_0\\
    \gamma_0\\
  \end{pmatrix}
  \label{eq:EV0}
\end{equation}
and
\begin{equation}
  \mathcal{H}_0= \hbar
  \begin{pmatrix}
    -\tfrac{1}{2}\omega_+'+\tfrac{3}{2}\omega &
    g_1  &   g_2 \\
    g_1  & \tfrac{1}{2}\omega_-'+\tfrac{1}{2}\omega & 0 \\
    g_2  & 0 & -\tfrac{1}{2}\omega_-'+\tfrac{1}{2}\omega  
  \end{pmatrix} .
  \label{eq:EVH0}
\end{equation}
This real symmetric matrix can be diagonalized. Three real eigenvalues $E_{0\xi}, \xi=1,2,3$
follow from the characteristic equation
\begin{equation}
  \det{(\mathcal{H}_0 - E_0  I)} = 0.
\end{equation}
The concomitant orthonormal eigenvectors $ \vec\eta_{0\xi}$ can then be calculated.
It results in the eigenstates in the $n=0$ subspace
\begin{equation}
  | E_{0\xi} \rangle  = \alpha_{0\xi} |1, \mathbf{0, 0} \rangle  + \beta_{0\xi} |0, \mathbf{1, 0} \rangle
  + \gamma_{0\xi} |0, \mathbf{0, 1} \rangle.
\end{equation}

The higher eigenvalues of $\mathcal N$ are equal to $n, n \ge 1$. Here we encounter a four-fold
degeneracy; the corresponding states are
   $ |n+1, \mathbf{0, 0} \rangle,  |n, \mathbf{1, 0} \rangle,  
    |n, \mathbf{0, 1} \rangle, |n-1, \mathbf{1, 1} \rangle.$
Consequently, we make the {\em Ansatz} for the eigenstates $|E_n \rangle$
\begin{widetext}
\begin{equation}
  | E_n \rangle = \alpha_n |n+1, \mathbf{0, 0} \rangle  + \beta_n |n, \mathbf{1, 0} \rangle 
   + \gamma_n |n, \mathbf{0, 1} \rangle  + \zeta_n |n-1, \mathbf{1, 1} \rangle.
\end{equation}
The eigenvalue equation $H|E_n\rangle = E_n |E_n\rangle$  explicitly yields for $n \ge 1$
\begin{eqnarray}
  && E_n\left\{\alpha_n |n+1, \mathbf{0, 0} \rangle  + \beta_n |n, \mathbf{1, 0} \rangle  
   + \gamma_n |n, \mathbf{0, 1} \rangle  + \zeta_n |n-1, \mathbf{1, 1} \rangle\right\}=  \\
   &&\hbar\alpha_n\left\{\left(-\tfrac{1}{2}\omega_+'+\omega(n+\tfrac{3}{2})\right)
  |n+1, \mathbf{0, 0} \rangle  + g_1\sqrt{n+1} |n, \mathbf{1, 0} \rangle  
+  g_2\sqrt{n+1} |n, \mathbf{0, 1} \rangle  \right\} \nonumber \\
&+&\hbar\beta_n\left\{\left(\tfrac{1}{2}\omega_-'+\omega(n+\tfrac{1}{2})\right)
  |n, \mathbf{1, 0} \rangle  +  g_1\sqrt{n+1} |n+1, \mathbf{0, 0} \rangle  
+  g_2\sqrt{n} |n-1, \mathbf{1, 1} \rangle  \right\} \nonumber \\
&+&\hbar\gamma_n\left\{\left(-\tfrac{1}{2}\omega_-'+\omega(n+\tfrac{1}{2})\right)
  |n, \mathbf{0, 1} \rangle  + g_1\sqrt{n} |n-1, \mathbf{1, 1} \rangle  
+  g_2\sqrt{n+1} |n+1, \mathbf{0, 0} \rangle  \right\} \nonumber \\
&+&\hbar\zeta_n\left\{\left(\tfrac{1}{2} \omega_+' +\omega(n-\tfrac{1}{2})\right)
  |n-1, \mathbf{1, 1} \rangle  +  g_1\sqrt{n} |n, \mathbf{0, 1} \rangle  
+  g_2\sqrt{n} |n, \mathbf{1, 0} \rangle  \right\}. \nonumber 
\end{eqnarray}
Taking the inner product with the basis states of
$\mathcal N$ leads to the matrix eigenvalue equations
\begin{equation}
   \mathcal{H}_n \vec \eta_n = E_n \vec \eta_n, \quad \text{where} \quad  \vec\eta_n = 
  \begin{pmatrix}
    \alpha_n \\
    \beta_n\\
    \gamma_n\\
    \zeta_n\\
  \end{pmatrix}
  \label{eq:EVn}
\end{equation}
and matrices
\begin{equation}
  \mathcal{H}_n= \hbar
  \begin{pmatrix}
    -\tfrac{1}{2}\omega_+'+\omega(n+\tfrac{3}{2}) &
    g_1\sqrt{n+1}  &   g_2\sqrt{n+1} & 0 \\
    g_1\sqrt{n+1}  & \tfrac{1}{2}\omega_-'+\omega(n+\tfrac{1}{2}) & 0 &
       g_2\sqrt{n}  \\
    g_2\sqrt{n+1}  & 0 & -\tfrac{1}{2}\omega_-'+\omega(n+\tfrac{1}{2}) & 
       g_1\sqrt{n}  \\
    0& g_2\sqrt{n}   & g_1\sqrt{n}  &
    \tfrac{1}{2}\omega_+'+\omega(n-\tfrac{1}{2}) & 
  \end{pmatrix}.
  \label{eq:EVHn}
\end{equation}
These real symmetric matrices can again be diagonalized and
four energy eigenvalues $E_{n\nu}, \nu=1,2,3,4$
are obtained for each $n$ by solving the characteristic equations
\begin{equation}
  \det{(\mathcal{H}_n - E_n  I)} = 0.
\end{equation}
As above the corresponding orthonormal eigenvectors $\vec \eta_{n\nu}$ 
can be computed. In this  way, we get the eigenstates the subspaces defined by
the value of $n$
\begin{eqnarray}
  | E_{n\nu} \rangle  &=& \alpha_{n\nu} |n+1, \mathbf{0, 0} \rangle  +
  \beta_{n\nu} |n, \mathbf{1, 0} \rangle  \nonumber \\
  &+& \gamma_{n\nu} |n, \mathbf{0, 1} \rangle  + \zeta_{n\nu} |n-1, \mathbf{1, 1} \rangle.
\end{eqnarray}
Note that, besides the RWA, no further approximations have been made in solving the
Schr\"odinger equation for the two-qubit JC Hamiltonian (\ref{eq:JCq2}).

\subsection{Numerical solution of the eigenvalue problems}
The eigenvalue problems defined by equations (\ref{eq:EV0}, \ref{eq:EVH0}, \ref{eq:EVn}, \ref{eq:EVHn})
will
be solved numerically for given values of the parameters. To this end, we use the 
Fortran 95 subroutine \texttt{jacobi} \cite{NR90} which does the job for real symmetric $n \times n$
matrices. It is based on Jacobi rotations, cf. \cite{Golub}.
The lowest energy levels for some typical parameter values are shown in Fig. \ref{fig:qubit2}.
Three energy scales are visible in the energy levels.  The energy differences between states
corresponding to a different eigenvalue of $\mathcal{N}$ are $\hbar \omega$. The larger
differences for equal eigenvalues of $\mathcal{N}$ are $\hbar(\omega-\tfrac{1}{2}(\omega_1'+\omega'_2))$
whereas the smallest ones are $\hbar(\omega'_1-\omega'_2)$. These relations are approximately
valid for small couplings.

\begin{figure}[htb]
\includegraphics[width=220pt] {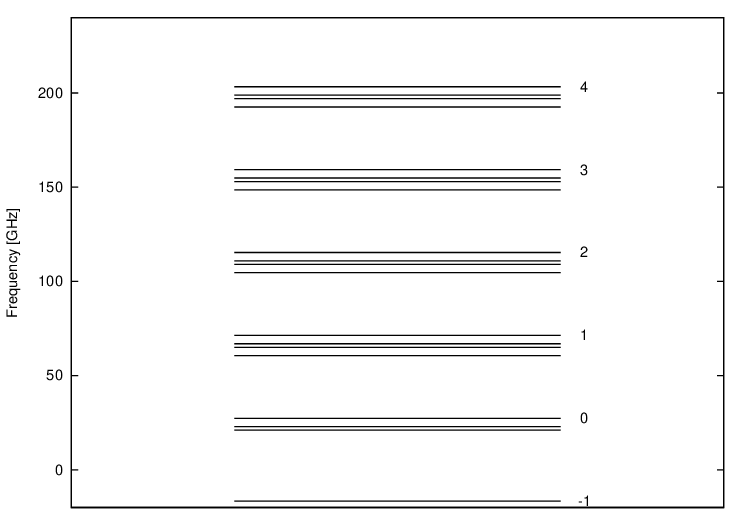}
\includegraphics[width=220pt] {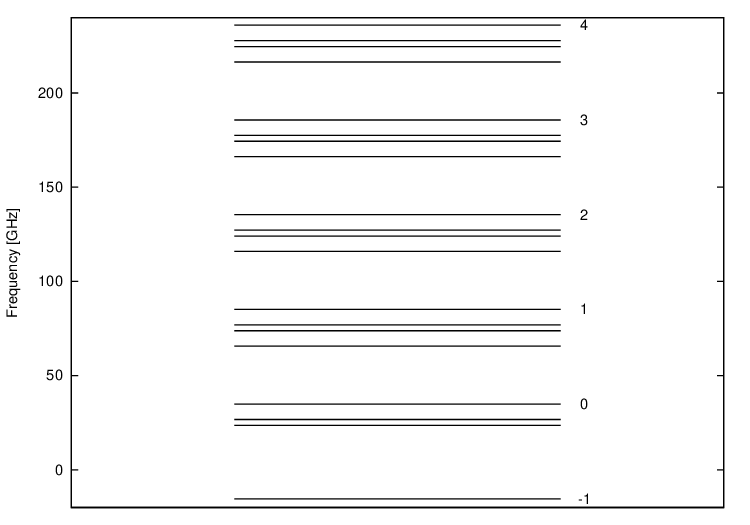}
\caption{\label{fig:qubit2} Energy levels $E/\hbar$
of the two-qubit Hamiltonian (\ref{eq:JCq2})
for parameter values in GHz;  $g_1/\hbar=0.1, g_2/\hbar=0.12$;
  l.h.s.: $f'_1=  6.0, f'_2=6.3,
  f= 7.0$; r.h.s.: $f'_1=  6.2, f'_2 = 6.7, f = 8.0$.}
\end{figure}

\end{widetext}

\section{Resonant coupling of $K$ qubits}\label{sec:Reso}
The system described above is extended to $K$ qubits with identical transition frequency $\omega'$.
It is the original Tavis-Cummings model for a single-mode quantized radiation field interacting
with $K$ molecules \cite{Tavis}. 
For resonant interactions a collective interaction which scales as $\sqrt{K}$
has been expected and observed \cite{Fink}.
Exploiting our formalism, we analyze such systems
 for $\omega=\omega'$ and in the
one-excitation subspace.

For the two-qubit system considered above, the results in the one-excitation subspace
yield the eigen energies
\begin{equation}
E=\tfrac{\hbar}{2}\omega \quad \text{and} \quad E_\pm = \tfrac{\hbar}{2}\omega \pm \hbar \sqrt{2}
\bar{g}_{12},
\end{equation}
with mean coupling 
$ \bar{g}_{12} = \sqrt{\tfrac{1}{2}(g_1^2+g_2^2)}$.
The corresponding eigenstates are respectively given by
\begin{eqnarray}
|E \rangle &=& \frac{1}{\sqrt{2}\bar g_{12}}\left(g_2 |0, \mathbf{1,0}\rangle
-g_1 |0, \mathbf{0,1}\rangle \right)  \\
|E_\pm \rangle &=& \pm \tfrac{1}{2}\sqrt{2} |1, \mathbf{0,0}\rangle 
+\frac{1}{2\bar g_{12}}\left(g_1 |0, \mathbf{1,0}\rangle
+g_2 |0, \mathbf{0,1}\rangle \right). \nonumber 
\end{eqnarray}
For obvious reasons $|E\rangle$ is referred to as dark state whereas $|E_\pm\rangle$
are called bright states in \cite{Fink}. 

We immediately proceed to the $K$-qubit case. In the one-excitation subspace the $(K+1) \times (K+1)$
Hamiltonian is explicitly given by
\begin{equation}
  \mathcal{H}_1= \hbar
  \begin{pmatrix}
    \Omega_K & g_1 & g_2 & \hdots &g_{K-1}  & g_K \\
    g_1  &   \Omega_K & 0 & \hdots & \hdots & 0 \\
    g_2 & 0 & \Omega_K & \hdots & \hdots & 0  \\
    \vdots & \vdots & \vdots &  \ddots & \vdots & \vdots \\
    g_{K-1} & 0 & \hdots & \hdots &  \ddots & 0 \\
    g_K  & 0 & \hdots & \hdots & 0 & \Omega_K 
  \end{pmatrix},
  \label{eq:EKOM}
\end{equation}
where $\Omega_K = \left(\frac{3-K}{2}\right)\omega$.
\begin{widetext}
We conjecture that the energy eigenvalues are
\begin{equation}
E=\hbar\Omega_K \quad \text{and} \quad E_\pm =  \hbar\Omega_K \pm \hbar \sqrt{K}
\bar{g}_{12\cdots K}
\end{equation}
with a $(K-1)$-fold degenerate dark subspace
\begin{equation}
|E \rangle = \alpha_1 |0, \mathbf{1,0, \cdots, 0}\rangle
 + \alpha_2 |0, \mathbf{0, 1, \cdots, 0}\rangle + \cdots
+\alpha_K |0, \mathbf{0, 0 \cdots, 1}\rangle
\end{equation}
and the bright states
\begin{equation}
|E_\pm \rangle = \pm \tfrac{1}{2}\sqrt{2} |1, \mathbf{0,0, \cdots, 0}\rangle 
+ \frac{1}{\sqrt{2K}\bar g_{12\cdots K}}\left(g_1 |0, \mathbf{1,0, \cdots, 0}\rangle
+g_2 |0, \mathbf{0,1, \cdots, 0}\rangle +\cdots
+g_K |0, \mathbf{0,0, \cdots, 1}\rangle \right).
\end{equation}
The mean coupling is defined as $\bar{g}_{12\cdots K} = \sqrt{\tfrac{1}{K}(g_1^2+g_2^2+\cdots+g_K^2)}$
and the coefficients $\alpha_k$ satisfy $\sum_{k=1}^K \alpha_k g_k = 0$.
These results can be verified by straightforward calculation of
$\mathcal{H}_1 |E \rangle$ and $\mathcal{H}_1 |E_\pm \rangle$. In this way we have confirmed  
the strength of the collective interaction to be 
$\bar{g}_{12\cdots K} \sqrt{K}$ 
as experimentally demonstrated for two and three qubits \cite{Fink}.
\end{widetext}

\section{Coupled qudit systems}\label{sec:Qudit}
The formalism developed thus far can be generalized to qudits, coupled by electromagnetic resonators.
A qudit is a finite-dimensional quantum system, obviously the generalization of qubits and qutrits. 
\subsection{Formalism}
We consider $K$ qudits and  allow qudits of different dimensionality
$D_k-1=2M_k$, with $M_k=1/2, 1, 3/2 \dots$.
The Hamiltonian of the coupled system is given in the RWA by
\begin{equation}
  H= \sum_{k=1}^K H_0^{[k]} + \sum_{p=1}^P H_{\text{res}}^{[p]}
  +\sum_{p=1}^P \sum_{k=1}^K H_{\text{int}}^{pk}.
\end{equation}
The qudit, resonator and interaction Hamiltonians are respectively given by
\begin{eqnarray}
  H_0^{[k]} &=& 
  \hbar \sum_{m=0}^{2M_k} \omega_{mk} |\bm{m}\rangle_k \langle \bm{m}|_k, \nonumber \\
  H_{\text{res}}^{[p]} &=&  \hbar  \omega_p(a_p^\dagger a_p+\tfrac{1}{2}), \nonumber \\
  H_{\text{int}}^{pk} &=&
  \hbar  \left(a_p^\dagger S_+^{[pk]}+a_pS_-^{[pk]}\right).
\end{eqnarray}
The `spin' raising/lowering operators $S_\pm^{[pk]}$
are explicitly defined as
\begin{eqnarray}
  S_+^{[pk]} &=& \sum_{l=0}^{2M_k-1} g_{ll+1}^{[pk]} |\bm l\rangle_k \langle \bm{l+1}|_k, \nonumber \\
  S_-^{[pk]} &=& \sum_{l=0}^{2M_k-1} g_{l+1l}^{[pk]} |\bm{l+1}\rangle_k \langle \bm l|_k.
\end{eqnarray}
Although they depend on the cavity index $p$ through the couplings $g_{ll+1}^{[pk]}$, as 
operators they only act on the qudit labelled by $k$. In typical applications like the transmon-cavity
model and a two-level atom-cavity system \cite{Mavro,Koch},
$g_{ll+1}\simeq g \sqrt{l+1}$, with dipole strength $g$. Here we only assume the couplings to be real
and symmetric which implies hermitian Hamiltonians.
Once again, we define the excitation-number operator for the system
\begin{equation}
  \mathcal{N} = \sum_{p=1}^{P} a_p^\dagger a_p 
  - \sum_{k=1}^{K} S^{[k]},
\end{equation}
with `spin' operators $S^{[k]}$
\begin{equation}
  S^{[k]} = \sum_{m=0}^{2M_k}(M_k-m) |\bm{m}\rangle_k \langle \bm{m}|_k.
\end{equation}
Their eigenstates and eigenvalues are given by
\begin{equation}
  S^{[k]} |\bm m\rangle_k = (M-m_k) |\bm m\rangle_k.
\end{equation}
It is also readily verified that for all $p,k$
\begin{equation}
  [S^{[k]}, S_\pm^{[ps]}] = \pm \delta_{ks} S_\pm^{[pk]}.
  \label{eq:C3}
\end{equation}
Analogously to the qubit-resonator system discussed above, it follows that the
excitation-number operator commutes with the Hamiltonian and is a conserved quantity.
Therefore, excitation-number operator and Hamiltonian are again simultaneously
diagonalizable.
The eigenstates of $\mathcal{N}$ are once more product states:
  $ |n_1,\dots,n_P, \bm{m_1},\dots,\bm{m_K} \rangle$.
The concomitant eigenvalues $N$  explicitly read 
\begin{equation}
  N = \sum_{p=1}^P n_p + \sum_{k=1}^K m_k - \sum_{k=1}^K M_k.
  \label{eq:eigqd}
\end{equation}
Re-introducing the notation $|N; l\rangle$ for the degenerate eigenstates of $\mathcal N$,
the matrix elements the Hamiltonian can be evaluated. First we get for 
the `free' Hamiltonians 
\begin{equation}
  \langle N'; l'| \sum_{k=1}^K H_0^{[k]} | N; l\rangle = 
   \hbar \delta_{N'N}\delta_{l'l} \sum_{k=1}^{K}  \omega_{mk}
\end{equation}
and
\begin{equation}
  \langle N'; l'| \sum_{p=1}^P H_{\text{res}}^{[p]} | N; l\rangle = 
    \hbar \delta_{N'N}\delta_{l'l} \sum_{p=1}^{P} \omega_p(n_p+\tfrac{1}{2}).
\end{equation}
We obtain for the interaction term
\begin{eqnarray}
  &&  \langle N'; l'| \sum_{p=1}^P \sum_{k=1}^K H_{\text{int}}^{pk} | N; l\rangle = \\
  &\hbar&\sum_{p=1}^P \sum_{k=1}^K \delta_{n'n}^{-p}\delta_{m'm}^{-k}
  \left(\delta_{n_p'n_{p+1}}\delta_{m_k'm_k-1}\sqrt{n_p+1}g_{m_k-1m_k}^{[pk]}\right. \nonumber \\
  &+& \left.\delta_{n_p'n_{p-1}}\delta_{m_k'm_k+1}\sqrt{n_p}g_{m_k+1m_k}^{[pk]}\right).\nonumber
\end{eqnarray}
Exploiting the various Kronecker deltas gives for nonvanishing matrix elements
$ N=  N' + n_p - n_p'+ m_k-m_k' =N$.
Thus it is shown that the Hamiltonian is diagonal with respect to $N$.
In these respective subspaces the matrices are symmetric as well. 
Consequently, the matrices have real eigenvalues and
- within their particular subspace- a complete set of eigenvectors. In other words,
they are diagonalizable. In this sense, the general multiple qudit-resonator problem is also solved.

\section{Two coupled qudits}\label{sec:Trans}
A relevant example of a qudit is the transmon in the three-level approximation. 
It has been shown \cite{Carlo} that one has to include the second excitation
level to achieve a two-qubit gate for two transmons. Here we explicitly consider two 
transmons as qutrits which are coupled by one resonator.
The Hamiltonian of this system can be written as
\begin{eqnarray}
  H &=& 
  \hbar \sum_{m=0}^{2} \omega_{m1} |\bm{m}\rangle_1 \langle \bm{m}|_1 +
  \hbar \sum_{l=0}^{2} \omega_{l2} |\bm{l}\rangle_2 \langle \bm{l}|_2 \nonumber \\
  &+&  \hbar  \omega(a^\dagger a+\tfrac{1}{2})  \\
  &+&\hbar  \left(a^\dagger S_+^{[1]}+aS_-^{[1]}\right)
  +\hbar  \left(a^\dagger S_+^{[2]}+aS_-^{[2]}\right). \nonumber
\end{eqnarray}
Approximating the transmons as anharmonic oscillators gives in terms of the charging energy
$E_C$ and the Josephson energy $E_J$ 
\begin{eqnarray}
  \hbar \omega_{01} &=& \tfrac{1}{2}\left(\sqrt{8E_J^{[1]}E_C^{[1]}}-E_C^{[1]}\right), \\ 
  \hbar \omega_{11}&=& 3 \hbar \omega_{01},
  \quad \hbar \omega_{21}= 5 \hbar \omega_{01} + \alpha^{[1]}. \nonumber
\end{eqnarray}
The anharmonicity reads $\alpha^{[1]}=-E_C^{[1]}$. For the second transmon analogous relations are valid.
The raising/lowering operators are defined as
\begin{eqnarray}
  S_+^{[1,2]} &=& \sum_{l=0}^{1} g_{ll+1}^{[1,2]} |\bm l\rangle_{1,2} \langle \bm{l+1}|_{1,2},
  \nonumber \\
  S_-^{[1,2]} &=& \sum_{l=0}^{1} g_{ll+1}^{[1,2]} |\bm{l+1}\rangle_{1,2} \langle \bm{l}|_{1,2}
\end{eqnarray}
and it is easily verified that
\begin{eqnarray}
 S_+^{[1,2]} | \bm 0 \rangle_{1,2} = 0, &\quad& 
  S_-^{[1,2]} | \bm 0 \rangle_{1,2} =  g_{10}^{[1,2]} | \bm 1 \rangle_{1,2}, \nonumber \\
  S_+^{[1,2]} | \bm 1 \rangle_{1,2} = g_{01}^{[1,2]} |\bm 0 \rangle_{1,2}, &\quad& 
  S_-^{[1,2]} | \bm 1 \rangle_{1,2} =  g_{21}^{[1,2]} | \bm 2 \rangle_{1,2}, \nonumber \\
  S_+^{[1,2]} | \bm 2 \rangle_{1,2} = g_{12}^{[1,2]} |\bm 1 \rangle_{1,2}, &\quad& 
  S_-^{[1,2]} | \bm 2 \rangle_{1,2} = 0.
\end{eqnarray}
The spin operators
$ S^{[1,2]}  = | \bm 0 \rangle_{1,2} \langle \bm 0 |_{1,2} - 
| \bm 2 \rangle_{1,2} \langle \bm 2 |_{1,2} $ 
are diagonal:
\begin{eqnarray}
  S^{[1,2]} | \bm 0 \rangle_{1,2} &=& |\bm 0 \rangle_{1,2}, \quad 
  S^{[1,2]} | \bm 1 \rangle_{1,2} =  0, \nonumber\\ 
  S^{[1,2]} | \bm 2 \rangle_{1,2} &=& -| \bm 2 \rangle_{1,2}. 
\end{eqnarray}
Herewith we again define the excitation number operator
\begin{equation}
\mathcal N = a^\dagger a - S^{[1]} -S^{[2]}.
\end{equation}
Its eigenstates are once more given by product states
\begin{eqnarray}
  \mathcal N |k, \mathbf{0, 0} \rangle  &=& (k-2) |k, \mathbf{0, 0} \rangle,   \nonumber \\
  \mathcal N |k, \mathbf{1, 0} \rangle  &=& (k-1) |k, \mathbf{1, 0} \rangle,   \nonumber \\
  \mathcal N |k, \mathbf{0, 1} \rangle  &=& (k-1) |k, \mathbf{0, 1} \rangle,   \nonumber \\
  \mathcal N |k, \mathbf{2, 0} \rangle  &=& k |k, \mathbf{2, 0} \rangle,   \nonumber \\
  \mathcal N |k, \mathbf{0, 2} \rangle  &=& k |k, \mathbf{0, 2} \rangle,   \nonumber \\
  \mathcal N |k, \mathbf{1, 1} \rangle  &=& k |k, \mathbf{1, 1}\rangle.
\end{eqnarray}
The state $|0, \bm 0, \bm 0 \rangle$  has the lowest eigenvalue of $\mathcal N$, {\it i.e.}, $N=-2$.
Evidently it corresponds to the ground state of the Hamiltonian
\begin{equation}
  H  |0, \mathbf{0, 0} \rangle  = \hbar \left(\omega_{01}+\omega_{02}+\tfrac{1}{2}\omega\right)
  |0, \mathbf{0, 0} \rangle.
\end{equation}
The first excitation subspace is three-fold degenerate with respect to the eigenvalues of $\mathcal{N}$
because the three states $|1, \mathbf{0, 0} \rangle, |0, \mathbf{1, 0} \rangle, |0, \mathbf{0, 1} \rangle$ 
have  $N =-1$. Hence we make the {\em Ansatz} for the eigenstates of $H$
\begin{equation}
  | E_{-1} \rangle  = \alpha |1, \mathbf{0, 0} \rangle  + \beta |0, \mathbf{1, 0} \rangle  
   + \gamma |0, \mathbf{0, 1} \rangle. 
\end{equation}
\begin{widetext}
The time-independent Schr\"odinger equation explicitly yields
\begin{eqnarray}
  &&\hbar\alpha\left\{\left(\omega_{01}+\omega_{02}+\tfrac{3}{2}\omega\right)
  |1, \mathbf{0, 0} \rangle  + g_{10}^{[1]} |0, \mathbf{1, 0} \rangle  
+  g_{10}^{[2]}|0, \mathbf{0, 1} \rangle  \right\} 
+\hbar\beta\left\{\left(\omega_{11}+\omega_{02}+\tfrac{1}{2}\omega\right)
|0, \mathbf{1, 0} \rangle  +  g_{01}^{[1]} |1, \mathbf{0, 0} \rangle  \right\} \nonumber \\
&+& \hbar\gamma\left\{\left(\omega_{01}+\omega_{12}+\tfrac{1}{2}\omega\right)
  |0, \mathbf{0, 1} \rangle 
+  g_{01}^{[2]} |1, \mathbf{0, 0} \rangle  \right\} 
= E_{-1}\left\{\alpha |1, \mathbf{0, 0} \rangle  + \beta |0, \mathbf{1, 0} \rangle  
   + \gamma |0, \mathbf{0, 1} \rangle \right\}.
\end{eqnarray}
Taking the inner product with the basis states of
of this subspace leads to the matrix eigenvalue equation
\begin{equation}
  \mathcal{H}_{-1} \vec \eta_{-1} = E_{-1} \vec \eta_{-1}, \quad \text{where} \quad  \vec\eta_{-1} = 
  \begin{pmatrix}
    \alpha \\
    \beta\\
    \gamma\\
  \end{pmatrix}
  \label{eq:tra-ex10}
\end{equation}
and
\begin{equation}
  \mathcal{H}_{-1}= \hbar
  \begin{pmatrix}
    \omega_{01}+\omega_{02}+\tfrac{3}{2}\omega &
    g_{01}^{[1]}  &   g_{01}^{[2]} \\
    g_{10}^{[1]}  & \omega_{11}+\omega_{02}+\tfrac{1}{2}\omega & 0 \\
    g_{10}^{[2]}  & 0 & \omega_{01}+\omega_{12}+\tfrac{1}{2}\omega  
  \end{pmatrix}.
  \label{eq:tra-ex1}
\end{equation}

The second excitation level corresponds to eigenvalue zero of $\mathcal N$. Its basis states 
follow analogously and lead to the {\em Ansatz} for the eigenstates of $H$:
\begin{equation}
  | E_0 \rangle = v_1 |2, \mathbf{0, 0} \rangle  + v_2 |1, \mathbf{1, 0} \rangle  
  + v_3 |1, \mathbf{0, 1} \rangle + v_4 |0, \mathbf{1, 1} \rangle  
  + v_5 |0, \mathbf{2, 0} \rangle  + v_6 |0, \mathbf{0, 2} \rangle  .
\label{eq:trans}
\end{equation}
The time-independent Schr\"odinger equation in the subspace can be rewritten as
matrix equation for the vectors $\vec v$ with components $v_l, l=1,\dots, 6$
\begin{equation}
  \mathcal{H}_0 \vec v = E_0 \vec v.
\end{equation}
The energy eigenvales are $E_0$ and the $6 \times 6$ matrix $\mathcal H_0$ reads
\begin{equation}
  \mathcal{H}_{0}= 
  \begin{pmatrix}
    \mathcal{H}_{11} & \mathcal{H}_{12} \\
    \mathcal{H}_{21} & \mathcal{H}_{12}  
  \end{pmatrix},
  \label{eq:tra-ex2}
\end{equation}
with the $3 \times 3$ matrices
\begin{eqnarray}
  \mathcal{H}_{11}&=& \hbar
  \begin{pmatrix}
    \omega_{01}+\omega_{02}+\tfrac{5}{2}\omega &
    \sqrt{2}g_{01}^{[1]}  & \sqrt{2}g_{01}^{[2]} \\
    \sqrt{2} g_{10}^{[1]}  & \omega_{11}+\omega_{02}+\tfrac{3}{2}\omega & 0 \\
    \sqrt{2} g_{10}^{[2]}  & 0 & \omega_{01}+\omega_{12}+\tfrac{3}{2}\omega  
  \end{pmatrix}, \nonumber \\
  \mathcal{H}_{22} &=& \hbar
  \begin{pmatrix}
     \omega_{11}+\omega_{12}+\tfrac{1}{2}\omega & 0 & 0 \\
      0 &  \omega_{21}+\omega_{02}+\tfrac{1}{2}\omega & 0 \\
     0 & 0 &  \omega_{01}+\omega_{22}+\tfrac{1}{2}\omega
  \end{pmatrix}
\end{eqnarray}
and
\begin{equation}
  \mathcal{H}_{12} = \hbar
  \begin{pmatrix}
     0 & 0 & 0 \\
     g_{01}^{[2]}& g_{12}^{[1]} & 0 \\
     g_{01}^{[1]} & 0 &g_{12}^{[2]}   
  \end{pmatrix}, \qquad
  \mathcal{H}_{21} = \hbar
  \begin{pmatrix}
    0 & g_{10}^{[2]} & g_{10}^{[1]}  \\
    0 & g_{21}^{[1]} & 0  \\
    0 & 0 & g_{21}^{[2]} 
  \end{pmatrix}.
\end{equation}

The program can of course be continued for higher excitation levels. The dimension of each subspace
is determined by its degeneracy with respect to the eigenvalues of the excitation number operator.
The eigenvalues and eigenstates follow by diagonalization of the respective subspace Hamiltonians.
Here we do not pursue this further but will explicitly solve the eigenvalue problems
for the first and second excitation level.

Some remarks are relevant for the numerical implementation. It is shown in \cite{Koch}
that one eventually can take the couplings real and symmetric.
As a consequence, the matrices (\ref{eq:tra-ex1}, \ref{eq:tra-ex2}) are symmetric
and  therefore can be diagonalized. The energy eigenvalues are real indeed, and the 
three/six eigenvectors are complete and orthogonal in the three/six dimensional subspaces.
Of course, they can be normalized to one.
Moreover, from \cite{Koch} it follows that
\begin{equation}
  g_{12}^{[1,2]} = \sqrt{2}g_{01}^{[1,2]}.
\end{equation}
The numerical values of all parameters are taken as in \cite{Carlo}, making a comparison
of results possible. 

Arguably the main result of \cite{Carlo} is the demonstration of 
a conditional phase gate. First, the one-excitation
spectrum is shown as a function of the frequency of one transmon, being varied
by tuning. An avoided crossing is visible in the spectrum. The resulting
interaction, however, is claimed to be too small for the applications.
We have re-calculated the one excitation spectrum using our formalism.
The result is depicted in 
Fig. \ref{fig:cross1} and is consistent with Fig. 1 in \cite{Carlo}.
The abovementioned C-phase gate relies on the two excitation spectrum,
presented as Fig. 2 in \cite{Carlo}.
Fig. \ref{fig:cross1} demonstrates that our formalism reproduces --at least qualitatively--
the spectrum and especially the avoided crossing as shown in \cite{Carlo}.

\begin{figure}[htb]
\includegraphics[width=220pt] {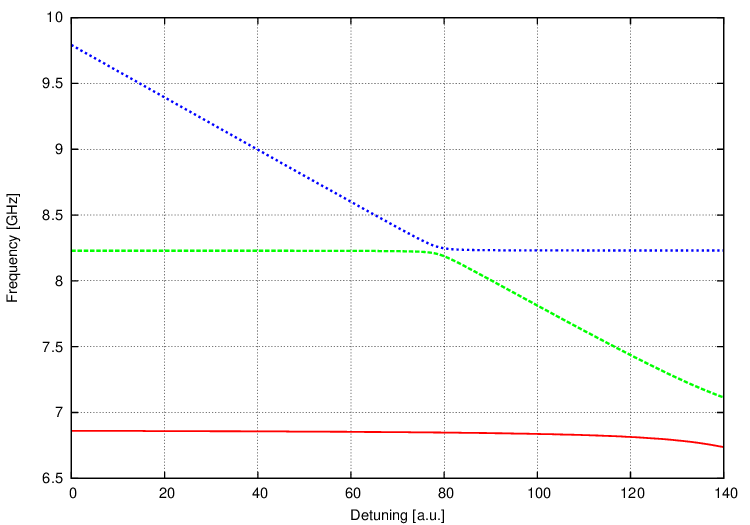}
\includegraphics[width=220pt] {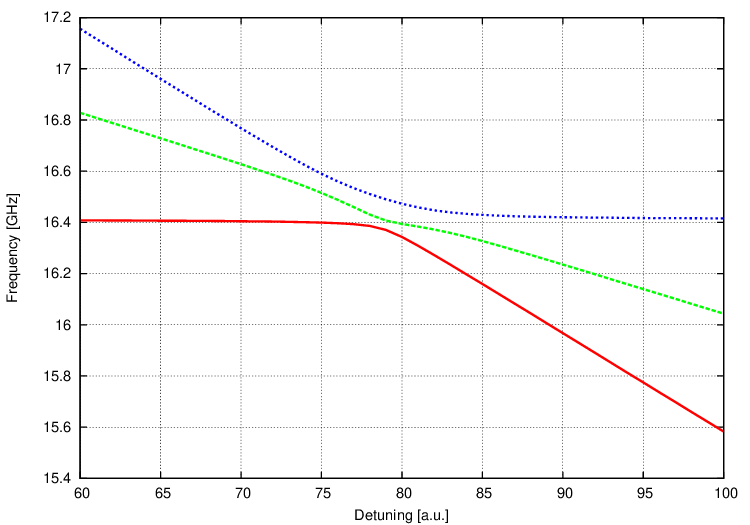}
\caption{\label{fig:cross1} Two coupled transmons, energy levels $E/\hbar$ as function
of the tuning of the frequency of one transmon.
L.h.s.  one-excitation subspace Hamiltonian (\ref{eq:tra-ex10}, \ref{eq:tra-ex1}), cf. Fig. 1 in \cite{Carlo};
  r.h.s two-excitation subspace Hamiltonian (\ref{eq:tra-ex2}), three highest energy levels, the corresponding states
have small coefficients $v_1, v_2, v_3$
in eq.(\ref{eq:trans}), cf. Fig. 2 in \cite{Carlo}.}
\end{figure}

\end{widetext}

\section{Time evolution}\label{sec:Time}
Finally, we address the evolution operator $U(t,t_0)$ of coupled multi-qudit systems.
In order to derive  formal expressions
we denote the eventually obtained eigenstates of the Hamiltonian as $ |E_{N\nu}\rangle $ where $N$ is the
excitation number and $\nu=1, \cdots, L(N)$; recall that $L(N)$ is the dimension of the
corresponding subspace. 
The evolution operator follows as 
\begin{eqnarray}
  U(t,t_0)=
  \sum_N\sum_{\nu}^{L(N)} e^{-iE_{N\nu}(t-t_0)/\hbar} 
  |E_{N\nu}\rangle \langle E_{N\nu}|,
  \label{eq:evo1}
\end{eqnarray}
cf. \cite{Cohen1}.
The energy eigenstates are linear combinations of the
eigenstates of $\mathcal N$
\begin{equation}
  |E_{N\nu}\rangle = \sum_{l=1}^{L(N)} c_{N\nu l} |N; l \rangle.
  \label{eq:enstate}
\end{equation}
The real coefficients $c_{Nl}$ and the energy eigenvalues $E_{N\nu}$
have been obtained from the diagonalization of the subspace matrices.
With (\ref{eq:enstate}) we rewrite the evolution operator (\ref{eq:evo1}) in terms of
these computed quantities
\begin{eqnarray}
  U(t,t_0)&=& \\
  \sum_N\sum_{\nu}^{L(N)} 
  \sum_{l,l'=1}^{L(N)} e^{-iE_{N\nu}(t-t_0)/\hbar} 
  &c_{N\nu l}&c_{N\nu l'} |N; l \rangle \langle N; l' | .  \nonumber
\end{eqnarray}
In practice the summation over $N$ has to be truncated.

\section{Summary and outlook}
This paper has provided a computational framework for the analysis of several
Jaynes-Tavis-Cummings systems in the RWA.
The latter approximation is assessed first for the simplest qubit-resonator system.
Generalizing the standard JC  Hamiltonians, the
formalism can deal with multiple qubits coupled to multiple resonators.
Eventually, the formal approach is extended to qudits, {\it i.e},
multi-level systems like spins and transmons in the qutrit
approximation \cite{Carlo}. The respective Hamiltionans are separated in decoupled subspace
Hamiltonians which correspond to fixed excitation numbers. This is possible since the
concomitant operator commutes with the Hamiltonian and is therefore conserved.
The symmetry of the Hamiltonian is of course directly related to the RWA.
In the respective subspaces, the corresponding symmetric matrices need to be diagonalized
in order to obtain the eigenstates, eigenvalues and evolution operators.
This is the purport of our, to the best of our knowledge new, semi-analytical formalism.
We have quantitatively demonstrated the framework at the hand of a
few examples. First, the system of two-qubits coupled to one resonator is analyzed completely.
The problem is extended to $K$ qubits with resonant couplings. In the one-excitation
subspace the  $\sqrt{K}$-scaling for 
the collective interaction strength between bright states as well as the appearance
of dark states \cite{Fink} has been derived. 
Finally, the coupled two-transmon system in the three-level approximation 
\cite{Carlo} has been examined.
Possible consequences for developments in quantum computing, in particular
for fault-tolerant quantum processors, the corresponding error corrections, stabilizers
and fidelities will be discussed in a subsequent paper \cite{NV2}. 

\section*{Acknowledgments}
The authors thank B. Criger and S. Poletto
for useful discussions and critical readings of the manuscript.
This research is supported by the Early Research Programme of the Netherlands
Organisation for Applied Scientific Research (TNO). Additional support from the
Top Sector High Tech Systems and Materials is highly appreciated.

\end{document}